\documentstyle[multicol,aps,prl,epsf]{revtex}

\begin{document}

\draft

\title{
Metal-Induced Gap States at Well Defined Alkali-Halide/Metal Interfaces
}

\author{Manabu Kiguchi$^1$, Ryotaro Arita$^2$, Genki Yoshikawa$^3$, 
Yoshiaki Tanida$^4$,
Masao Katayama$^3$, Koichiro Saiki$^{1,3}$, Atsushi Koma$^3$ and Hideo Aoki$^2$
}

\address{$^1$Department of Complexity Science $\&$ Engineering, Graduate 
School of Frontier Sciences, The University of Tokyo, Hongo, Tokyo 
113-0033, Japan}
\address{$^2$Department of Physics, Graduate School of Science, 
The University of Tokyo, Hongo, Tokyo 
113-0033, Japan}
\address{$^3$Department of Chemistry, Graduate School of Science, 
The University of Tokyo, Hongo, Tokyo 
113-0033, Japan}
\address{$^4$Fujitsu Laboratories Ltd., Atsugi, Kanagawa 243-0197, Japan}

\date{\today}

\maketitle

\begin{abstract}
In order to search for states specific to insulator/metal interfaces,
we have studied epitaxially grown interfaces with 
element-selective near edge X-ray absorption fine structure (NEXAFS). 
An extra peak is observed below the bulk edge onset for 
LiCl films on Cu and Ag substrates. The nature of chemical 
bonds as probed by X-ray photoemission spectroscopy and Auger electron
 spectroscopy remains unchanged, so we regard this as
 evidence for metal-induced gap states(MIGS) formed by the proximity
 to a metal, rather than local bonds at the interface.  The 
dependence on the film thickness shows that the MIGS are as thin as 
one monolayer.  An {\it ab
 initio} electronic structure calculation supports the existence of the
 MIGS that are strongly localized at the interface.
\end{abstract}

\medskip

\pacs{PACS numbers: 73.20.-r, 73.40.Ns, 71.15Mb}

\begin{multicols}{2}
\narrowtext
{\it Introduction} 
While there is mounting interest in the nature of the
``heterointerface" (solid-solid interfaces between very dissimilar
materials), insulator/metal interfaces are especially intriguing, since
they provide fascinating possibilities such as metal-insulator
transition\cite{jpc4,Anderson}, band gap narrowing \cite{prb64} and
superconductivity \cite{Tc} as well as technological ones such as 
catalysis, magnetic tunneling junctions, etc. Despite the interest,
electronic structures characteristic of the insulator/metal interface
have not been studied satisfactorily, for good
reasons. First, well-defined interfaces are hard to prepare due to the
different nature of chemical bonds. 
Second, signals from the interface are obscured by
significant contribution from the substrate in conventional 
experimental methods such as
ultraviolet photoemission spectroscopy(UPS), inverse photoemission
spectroscopy, or electron energy loss spectroscopy(EELS).
  
Recently, an interesting experimental result was reported for
insulator/metal systems. Muller {\it et al.} studied \{222\}MgO/Cu
interfaces in fine particles (rather than in a film) with
transmission EELS, and succeeded in observing 
metal-induced gap states (MIGS) for the first time\cite{prl80}. 
The concept of MIGS was first introduced for
semiconductor/metal junctions in discussing the Schottky barrier at the
interface\cite{prl52}, and subsequently applied to insulator/metal
interfaces\cite{jpc8}. MIGS are thought to accompany metal wave
functions whose (exponential) tails penetrate into the insulating side
of the interface. Schintke {\it et al.} studied MgO/Ag(001) with scanning
tunneling spectroscopy and found a state in the band gap\cite{prl87}. 
However, the first point raised above remains: Since \{222\}MgO is a polar 
(hence presumably metallic) surface, whether the pre-peak observed for 
\{222\}MgO/Cu originates from the polar surface or from the MIGS has not been definite, although their study is pioneering.  Furthermore, \{222\} MgO/Cu is not a well-defined interface due to a large lattice mismatch. For MgO/Ag(001), the strong
hybridization between the O 2p band and the Ag 5sp band is expected 
to dominate the interface\cite{prb59}. 
All these should obscure the identification of the MIGS in its proper sense, i.e., states formed solely by the proximity to a metal rather than by local bonds.

In order to clarify this, we propose here to prepare well-defined 
interfaces by exploiting 
our recent studies, in which we have revealed that some alkali halides grow
heteroepitaxially on metal substrates in a layer-by-layer fashion, so
 we end up with {\it atomically well-defined insulator/metal interfaces}
 \cite{prb63,prb66,saiki}. Being epitaxially grown, the 
 number of atomic layers can also be controlled, which helps to
 probe the nature of the interface state. In a previous study we
 have examined the epitaxially grown LiCl/Cu(001), but
 were unable to find an indication for the
 interface state\cite{ss522}, presumably because the methods employed 
 (UPS and EELS) pick up signals from the substrate. 

Here we adopt near edge x-ray
 absorption fine structure (NEXAFS) to study electronic
 structures at the LiCl/metal interfaces, where the second point raised
 above comes in. Namely, we adopt Cl $K$-edge NEXAFS,
 which is based on X-ray absorption by Cl atoms, and hence
 provides information on the LiCl film with negligible influences of the substrate. Furthermore, NEXAFS, with its high-energy photons,
 can probe very deep interfaces, so suited for obtaining the dependence of the 
interface states on the thickness of the insulating layer.

The results obtained here for LiCl thin 
films on Cu(001) and Ag(001) indeed exhibit clear evidence for the 
electronic states
 intrinsic to the insulator/metal interface which are as thin as one
 monolayer(ML)[1 ML$=2.6$ \AA]. We have then compared the 
 experimental result with an {\it ab-initio} density
 functional calculation, which supports
 the existence of the MIGS that are strongly localized at the interface. 

{\it Experimental}
 Epitaxial LiCl films were grown on metal substrates at 
300 K using a Knudsen cell. Cl-$K$ edge
 NEXAFS was carried out at the soft X-ray double-crystal
 monochromator station BL-11B in the Photon Factory in Institute of Materials
 Structure Science\cite{ohta}. The Cl-$K$ NEXAFS
 provides information on the unoccupied Cl-p states\cite{commentcorehole}. 

\begin{figure}
\begin{center}
\leavevmode\epsfysize=45mm \epsfbox{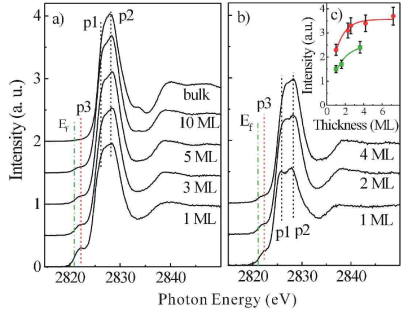}
\caption{The Cl-$K$ edge NEXAFS spectra in LiCl films grown on (a)
 Cu(001) and (b) Ag(001) for various thicknesses of the LiCl layer. 
E$_{F}$ indicates the position of the Fermi level as determined from XPS. 
The inset (c) shows the intensity of
 the pre-peak (not normalized by the edge-jump unlike in the main panel)
 versus the film thickness, where the curves are the least-square fit. 
}
\label{fig1}
\end{center}
\end{figure}

 Figure~\ref{fig1} shows the Cl-$K$ edge NEXAFS spectra for
 LiCl/Cu(001) and LiCl/Ag(001) taken at grazing X-ray incidence(15$^{\circ}$) for
 various thicknesses of the LiCl layer. All the spectra are normalized
 by their edge-jumps.
Two NEXAFS peaks are observed at 2827 eV (p1) and 2829 eV (p2) for bulk LiCl. 
 Now, a new finding here is a pronounced pre-peak (p3) appears just
 below the bulk edge onset, which is clearly visible 
 for thinner LiCl layers. Hereafter we will focus on 
 the pre-peak.
 
 There are two points to note. First, the pre-peak exists even
 for a 10 ML LiCl on Cu(001). Namely, although the peak may
 seem more prominent with decreasing the film thickness in Fig.~\ref{fig1},
 this is an artifact of the normalizing the curve by the edge-jump. 
 As we shall elaborate below in Fig.~\ref{fig1}(c), 
 the pre-peak appears at 1ML and its intensity rapidly saturates. 
 Second, the pre-peak is observed for both LiCl/Cu(001) and LiCl/Ag(001). For
 LiCl/Cu(001), LiCl grows with the [100] axis rotated by 45$^{\circ}$
 away from that of the Cu substrate, where both Li and Cl
 atoms sit on the fourfold hollow sites due to matched 
Li-Cl and Cu--Cu distances\cite{prb63,prb66}.  
On the other hand, LiCl on Ag(001) grows without an
 azimuthal rotation, for which a commensurate structure becomes
 impossible \cite{prb66}. Despite the difference in the interface
 structure between Cu and Ag, the pre-peak is observed in both
 systems, which precludes local
 structures at the interface as the origin of the pre-peak.

\begin{figure}
\begin{center}
\leavevmode\epsfysize=30mm \epsfbox{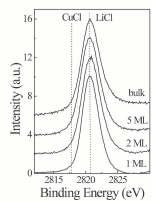}
\caption{The Cl 1s XPS spectra in LiCl/Cu(001) for various thicknesses
 of the LiCl layer with a photon energy of 2900 eV. The binding energies
 are defined with respect to the Fermi level, where the
 binding energy of the bulk CuCl is also indicated.}

\label{fig2}
\end{center}
\end{figure}

 We next confirm whether the nature of chemical bonds are modified around
 the interface by making use of X-ray photoemission spectroscopy
 (XPS) and Auger electron spectroscopy (AES). For LiCl/Cu(001), the Cu MMM
 AES peak does not shift from that of the clean Cu(001) within 0.1 eV,
 which implies that the interfacial Cu atoms remain almost neutral\cite{jvs12}. 
Figure ~\ref{fig2} shows the Cl 1s XPS spectra for LiCl/Cu(001)
 for various thicknesses of LiCl. For comparison, the spectrum for a bulk
 LiCl crystal and the peak position of a bulk CuCl crystal
 are also measured, which shows that the Cl 1s
 binding energy in LiCl is larger by 2 eV than that of CuCl. If the LiCl
 film interacted with the Cu substrate to form new chemical bonds at
 the interface, a satellite peak at the lower binding energy side or the
 broadening of the peak would arise.  However, both the binding energy and the
 width of the Cl 1s peak do not change with the film thickness, nor
 does a satellite peak appear in LiCl/Cu(001). These show that
 the chemical state of Cl in LiCl/Cu(001) is virtually the same as that
 of bulk LiCl with no chemical bonds formed at the interface.
 The similar results are obtained from Cl 1s XPS spectra for LiCl/Ag(001).
 
 Given the above NEXAFS, XPS,
 and AES results indicating that the pre-peak originates not from chemical
 bonds at the LiCl/metal interface, the gap states arise from the 
 proximity to a metal, so we can regard them as MIGS \cite{ss522,ohta}.
 We can also note that the position of the Fermi
 level, which can be estimated from the binding energy of Cl 1s XPS spectra as
 indicated in Fig.~\ref{fig1}, is seen to be located
 right in the pre-peak.  So a finite density of states at E$_{F}$ implies
 that LiCl is {\it metallized} at the interface.

 Let us discuss the character of the MIGS in more
 detail from the NEXAFS data.  
 First, we can exploit the epitaxy-grown samples to estimate the
 decay length of the MIGS into the insulating side from the dependence
 of the intensity of the pre-peak on the thickness of the insulating
 layer. Since the probing depth of NEXAFS is orders of magnitude
 greater (typically $\geq$ 10000 \AA) than the atomic scale, 
 NEXAFS intensity should directly reflect the decay of the MIGS 
 into the insulator.  If we assume an exponential decay to fit the
 experimental data, we end up with the decay length 
$2.6\pm 0.3$ \AA {} for LiCl/Cu(001) and $2.9\pm 0.7$ \AA {} 
for LiCl/Ag(001) (Fig.~\ref{fig1}(c)).  
So the MIGS are indeed localized 
within a few \AA ($\sim$ as small as one monolayer) 
of the interface. 

\begin{figure}
\begin{center}
\leavevmode\epsfysize=30mm \epsfbox{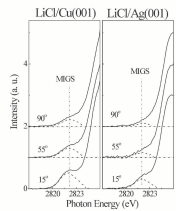}
\caption{Polarization dependence of the Cl-$K$ edge NEXAFS spectra for 1
 ML LiCl/Cu(001) and 1 ML LiCl/Ag(001). 
 The MIGS-derived peaks (difference between the spectra for 
 bulk LiCl and LiCl/Cu) are indicated by dotted lines.}

\label{fig3}
\end{center}
\end{figure}

Second, we have probed the geometry of the MIGS wave function 
from the polarization dependence of NEXAFS.  
The pre-peak has turned out to be greater for grazing X-ray incidence 
(Fig.~\ref{fig3}), which clearly shows that the 
MIGS are p$_z (z \perp$ surface)-like\cite{nex}.

{\it Ab-initio calculaton} 
Formation of MIGS with a thickness $\sim$ 1 ML is theoretically 
intriguing, since a simple model for the interface 
(e.g., tight-binding insulator/jellium) would predict a 
negligible ($\ll 1$ ML) penetration (for the gap of the insulator 
$\sim$ several eV) and would completely fail.  
So we have performed a 
 first-principles calculation based on the local density functional
 theory (LDF) with pseudopotentials and plane-wave basis.  
 We have adopted two models, one for LiCl/Cu(001)
 and the other for LiCl/jellium, to examine whether the MIGS
 come from the specific metallic substrate (copper here) or a general
 property of the proximity to an electron gas. The calculation 
 for LiCl/Cu is 
 performed on periodically repeated slabs with five Cu layers sandwiched
 from top and bottom by one LiCl layer each 
with the slabs separated by a vacuum of height 50 \AA.  
The in-plane unit cell contains only two Cu atoms and one LiCl, 
 which is a virtue of the commensurate structure. 

\begin{figure}
\begin{center}
\leavevmode\epsfysize=45mm \epsfbox{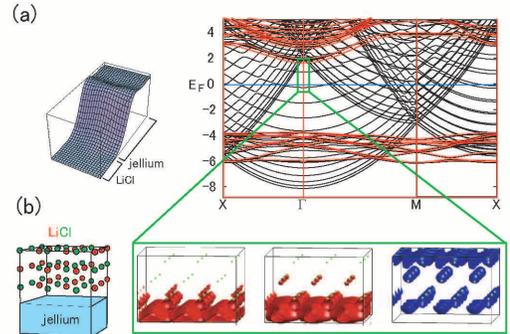}
\caption{(a) The band structure of 1 ML LiCl/Cu(001) (black) as compared with 
that for an isolated 1 ML LiCl (red).  
 (b) Contours of the
 absolute value of the LDA wave functions for the in-gap states
 (enclosed by a green square), along with the atomic
 configuration.  The wave functions having amplitudes on both Cl and
 Cu atoms are shown in red. Top left inset shows 
the charge distribution
 difference between LiCl/Cu(001) and an isolated LiCl.}
\label{fig4}
\end{center}
\end{figure}

 Figure~\ref{fig4} (a) shows the band structure of 1 ML
 LiCl/Cu(001). For comparison we have also obtained the band structure
 of an isolated 1 ML LiCl film without the metallic substrate.  
 The isolated LiCl has a large 
 band gap, although the size of the gap is, as
 usually the case with LDF, underestimated by 2-3 eV. When the 1 ML
 LiCl is put on Cu(001), new bands appear in the gap. We have 
 characterized their nature by looking at the amplitude of the LDF wave
 functions. 

If we look at the three states just above E$_{F}$ 
 at $\Gamma$ point,
 the two in-gap bands closest to E$_{F}$ have indeed
 amplitudes that are localized at the interfacial Cl and Cu atoms 
 (Fig.~\ref{fig4} (b)). Existence of such states is, theoretically, 
 remarkable, given the fact that LiCl is a very good insulator.   
They are p$_{z}$-like, in agreement 
 with the polarization dependence seen in NEXAFS.
So we identify that
 the MIGS observed by NEXAFS should correspond to these states. 

 Another theoretical observation is that the new state is by no means
 a result of a charge transfer across Cu and LiCl. This can be seen in
 the inset of Fig.~\ref{fig4}, which displays the difference in the
 charge density between LiCl/Cu(001) and the isolated LiCl film. If
 there were a charge transfer, the charge difference would have an
 amplitude on the LiCl side as well, while the result shows that it is
 entirely confined to the Cu side. The absence of charge transfer
 can explain the absence of core level shifts at the
 interface in Cl 1s XPS result(Fig.~\ref{fig2}).

\begin{figure}
\begin{center}
\leavevmode\epsfysize=45mm \epsfbox{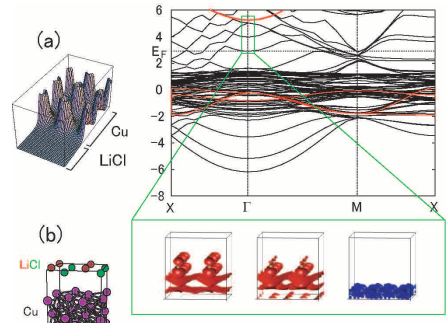}
\caption{a) The band structure of 3 ML LiCl/jellium 
with $r_{s}$=2.5 (black) 
as compared with that for an isolated 3 ML LiCl (red). 
(b) Contours of the absolute value of the LDA
 wave functions), along with the atomic
 configuration. The wave functions having amplitudes on both Cl and
 jellium are shown in red. Top left inset shows 
the charge distribution difference between LiCl/jellium
 and an isolated LiCl.}

\label{fig5}
\end{center}
\end{figure}

 In order to theoretically estimate the decay length of the MIGS into the
 insulator, we have then performed an LDF calculation for 3 ML
 LiCl attached to a jellium of thickness 80 \AA (from top and
 bottom). The replacement of the Cu substrate with a jellium also 
enables us to examine
 whether a simplest possible model for the electron gas would already
 exhibit MIGS. Figure~\ref{fig5}(a) shows the band structure of 3 ML
 LiCl/jellium, where the jellium has $r_{s}$=2.5. We have again new
 in-gap states whose wave
 functions (Fig.~\ref{fig5}(b)) just extend to the interfacial 
 (i.e., the first -layer from the
 jellium) Cl atoms. This result for LiCl/jellium endorses 
 that (i) MIGS are characteristic of insulator/metal interfaces rather than
 of local chemical bonds, and 
 (ii) $\sim$ 1ML penetration depth of the MIGS is rather general.

 To systematically confirm how the MIGS appear, we 
 have then varied the density of electrons in the metallic side 
by varying $r_s$, which is the sole parameter characterizing the jellium.  
 In the LDF result shows that MIGS appear 
 with its penetration depth only 
 weakly dependent on $r_s$. This accounts for the experimental result
 that MIGS are observed for both Cu(with $r_s=2.7$) and Ag($r_s=3.0$) 
 substrates. 

 Finally, let us point out that the electronic structure specific to
 insulator/metal interfaces can possibly have interesting implications
 for superconductivity. Discussions on superconductivity at interfaces
 have a long history. A well-known proposal is by Ginzburg {\it et al.}
 \cite{Tc} for the possibility of superconductivity by exciton mechanism
 at metal/semiconductor interfaces.  The presence of excitons
 requires a wide-gap insulator, which will, however, prevent the metallic
 carriers to penetrate into the insulator, so that 
 the coexistence (in real
 space) of excitons and metallic carriers are difficult. To circumvent
 this they had to consider a hypothetical 
 density of states that has two closely separated peaks in a wide band
 gap, where the former are required to enable the carriers in the
 metallic side to easily tunnel into the semiconductor side.

 By contrast, in the present MIGS in 
 insulator/metal interfaces 
 we automatically have a coexistence of excitations (associated with the 
 wide band gap ($\simeq 9$ eV for LiCl) of the insulator) 
 and carriers. To realize a high $T_C$ we have further to realize 
strong interaction between
 the carriers and excitons, which should also 
be fulfilled due to the penetration of MIGS into the
 insulating layer. So we may envisage that the insulator/metal interface
 as discussed here may provide a possible ground for
 superconductivity. 
 
This work was supported by a Grant-in-Aid for Scientific
 Research and Special Coordination Fund 
 from the Ministry of Education of Japan.  (Creative
 Scientific Research Project, No. 14GS0207).  The present work was performed under the approval of Photon Factory Program Advisory Committee. The numerical calculation was performed with
SR8000 in ISSP, University of Tokyo.

\end{multicols}
\end{document}